\documentclass[aps,floatfix,twocolumn,a4paper,showpacs, nofootinbib,
superscriptaddress,10pt]{revtex4}
\usepackage{graphicx,float}\usepackage{graphicx,float}
\usepackage[all]{xy}
\usepackage{amsmath,upgreek}
\usepackage{amssymb}
\usepackage{color}
\usepackage{epsfig}		
\usepackage{graphicx,epstopdf}
\usepackage{subfigure}
\usepackage{pdfpages}
\usepackage[colorlinks,hyperindex]{hyperref}

\definecolor{green1}{RGB}{0,128,0} 
\hypersetup{hidelinks,backref=true,pagebackref=true,hyperindex=true,colorlinks=true,breaklinks=true,urlcolor= blue}
\hypersetup{%
  colorlinks = true,
  linkcolor  = blue,
  citecolor = cyan,
}
\usepackage{bookmark,textgreek}
\usepackage{hyperref,color,xcolor}
\hypersetup{hidelinks,hyperindex=true,colorlinks=true,breaklinks=true,urlcolor= blue}
\hypersetup{%
  colorlinks = true,
  linkcolor  = blue
}

\newcommand{\bes}{\begin{subequations}}
\newcommand{\ees}{\end{subequations}}
\def\ben{\begin{eqnarray}}
 \newcommand{\bltx}{\textcolor{black}}
\def\een{\end{eqnarray}}
\def\be{\begin{equation}}
\def\ee{\end{equation}}

\begin{document}

\title{Tensor mesons, AdS/QCD and information}

\author{Luiz F.  Ferreira  }\email{luizfaulhaber@if.ufrj.br}
\affiliation{Federal University of ABC, Center of Mathematics,  Santo Andr\'e, Brazil}
\affiliation{Federal University of ABC, Center of Physics,  Santo Andr\'e, Brazil.}
\author{R. da Rocha}
\email{roldao.rocha@ufabc.edu.br}
\affiliation{Federal University of ABC, Center of Mathematics, Santo Andr\'e, Brazil}
\affiliation{HECAP, International Centre for Theoretical Physics (ICTP), Strada Costiera 11, 34151 Trieste, Italy}

\begin{abstract}
The Kaluza-Klein tower of higher spin-$S$ tensor meson resonances is here scrutinized in the AdS/QCD hard wall model, encompassing the already established resonances  
$\rho(770)$, $f_2(1270)$, $\omega_3(1670)$, $f_4(2050)$, $\rho_5(2350)$, $f_6(2510)$ in PDG.  A hybrid model employs both information theory and AdS/QCD, where  configurational-entropic Regge trajectories, relating the configurational entropy of the tensor mesons family to their $S$ spin, and also to their experimental mass spectra, are derived and analyzed. Therefore, the mass spectra of higher spin-$S$ tensor meson resonances is obtained and compared to the existing data in PDG.  
 \end{abstract}
\maketitle

\section{Introduction}

\bltx{The concept of configurational entropy (CE) designates the measure of spatial correlations in any physical system. The CE is interpreted as a lossless compression rate limit of information, representing the minimum quantity of bits that can encrypt a message. No code takes less number of bits per symbol than the CE source, which consists of the informational measure of  complexity of
a system that has spatial profile, with respect to the system modes in the momentum space \cite{Shannon:1948zz,Gleiser:2011di,Gleiser:2012tu}.
Hence,  the lower the information entropy, the lower the information encoded in the frequency modes is required to represent 
the shape of the physical system. When a system is totally predictable, namely, the distribution of probability  is unity for a particular symbol, and vanishes for all the others, then the CE vanishes. 
Localized solutions of PDEs are examples of non-trivial spatial complexity, as well as  the correlation of thermal fluctuations in a system undergoing continuous phase transitions. 
 Modes in a system corresponding to hadronic states with lower CE 
 have been shown to be more dominant and configurationally stable, being more detectable and observable, from the experimental point of view   \cite{Bernardini:2016hvx}. } 
 \bltx{The CE defines the best lossless compression of any  communication inside a physical system. The higher the CE, the more unpredictable and configurationally unstable the system is. The expression for the CE is similar to the Boltzmann--Gibbs entropy, measuring the degree of disorder in a physical
system. Hence, the CE is also a measure how much information is needed to determine a microstate. In this way, thermodynamic entropy  is proportional to the quantity of Shannon information that is required to define the microscopic state of a given system. A thoroughly exposition of the CE paradigm, even in its  quantum version  and its relationship to the quantum von Neumann entropy are detailedly exposed by Witten \cite{Witten:2018zva}.}

\bltx{The CE has been  evidenced to be a prominent tool to study QCD, specially in regarding AdS/QCD, endowing QCD with new methods analyze and scrutinize  physical phenomena. Indeed, the CE has been playing prominent roles in both the hard and soft wall  AdS/QCD models, further studying new features of QCD and corroborating to data in PDG. Hadronic states with lower CE either require less energy to be yielded, or are more frequently observed as more dominant hadronic states, than their configurationally unstable counterparts, or even both. 
The recently introduced concept of configurational-entropic Regge trajectories have been employed to study and also to predict the mass spectra of the next generation of higher spin resonances of four light-flavour meson families, $a_1$, $f_0$ and $\rho$   \cite{Bernardini:2018uuy,Ferreira:2019inu}. Formerly,  other features of mesonic resonances have been already studied under the CE paradigm \cite{Bernardini:2016hvx}. Besides, the CE has been employed to study the configurational stability of scalar glueballs in  AdS/QCD \cite{Bernardini:2016qit}. Ref. \cite{Braga:2017fsb} used the CE to discuss quarkonia production at zero temperature, whereas a more intricate development, using the CE, approached the finite temperature case of bottomonium and charmonium resonances  \cite{Braga:2018fyc}. 
Besides, Ref. \cite{Ferreira:2020iry} derived, using AdS/QCD, the mass spectra of the next generation of $J^P=\frac12^+$ and $J^P=\frac32^-$ nucleon and higher spin baryon resonances both at zero and at finite temperatures, and  them compared to already established baryonic states in PDG. }
 Light- and heavy-flavour mesonic excitations in chiral condensates, glueball resonances, baryons and the quark-gluon plasma have been also investigated 
under various aspects of the CE that underlies QCD \cite{Barbosa-Cendejas:2018mng,Colangelo:2018mrt,daSilva:2017jay,Ma:2018wtw}.  Other applications of the CE in QCD and in the standard model can be found in Refs. \cite{Karapetyan:2016fai,Karapetyan:2017edu,Karapetyan:2018oye,Karapetyan:2018yhm,Karapetyan:2019fst,Alves:2014ksa}.  Besides, the CE was shown to be an appropriate paradigm to study phase transitions as CE critical points \cite{Gleiser:2014ipa,Gleiser:2018kbq,Sowinski:2017hdw,Sowinski:2015cfa,Gleiser:2015rwa,roldao,Bazeia:2018uyg,Correa:2016pgr}, including the stability of stellar distributions \cite{Casadio:2016aum,Braga:2016wzx,Lee:2018zmp,Fernandes-Silva:2019fez}. 
  
 {\color{black}{The AdS/QCD approach has proved to be an important tool to investigate non-perturbative aspects of QCD as confinement \cite{
EKSS2005,Katz:2005ir,Karch:2006pv,Colangelo:2008us,Zhang:2010tk,Colangelo:2011sr,Brodsky:2014yha}. The construction of  holographic models for QCD-like theories is implemented by considering deformations in the AdS/CFT correspondence \cite{Maldacena:1997re,Witten:1998qj,gub}. For instance, the hard wall AdS/QCD model \cite{Polchinski:2001tt,BoschiFilho:2002ta,BoschiFilho:2002vd}  consists of introducing an infrared (IR) cut-off, which corresponds to an IR mass in the gauge theory, into the AdS space in order to break conformal invariance. In the case of the soft wall model \cite{Karch:2006pv}, the conformal invariance is broken  introducing a dilaton field that acts as a smooth IR cut-off. 
 }}

Our main aim here is to study the $\rho(770)$, $f_2(1270)$, $\omega_3(1670)$, $f_4(2050)$, $\rho_5(2350)$, $f_6(2510)$ family of tensor mesons in PDG \cite{pdg1}, in the hard wall AdS/QCD, under the CE.  The hard wall AdS/QCD can derive their masses with a good experimental accuracy up to spin $S=5$ \cite{Katz:2005ir}. However for $S\geq 6$ there is already a gap of 14.6\% between the predicted and the experimental mass of the $f_6(2510)$ tensor meson resonance. To derive a more trustworthy mass spectra for the higher spin-$S$ resonances in the tensor meson family, for $S=7, 8, 9$, the information 
content of AdS/QCD will be explored, using the CE together with the experimental masses of the already observed tensor mesons in PDG \cite{pdg1}.
As neither the soft wall nor the hard wall AdS/QCD model match experimental data  corresponding to heavier Kaluza-Klein (KK) resonances, we focus our analysis in the lightest KK excitations, corresponding to the $\rho(770)$, $f_2(1270)$, $\omega_3(1670)$, $f_4(2050)$, $\rho_5(2350)$, $f_6(2510)$ family of tensor mesons in PDG \cite{pdg1}.  
Therefore, equipped with the apparatus involving the CE that underlies the hard wall AdS/QCD model, 
 configurational-entropic Regge trajectories, for the tensor mesons family, will be derived and scrutinized. Here tensor meson resonances up to spin $S=9$ are analyzed. Two types of configurational-entropic Regge trajectories to be studied, relating the tensor meson family CE both to the spin-$S$ tensor resonances and their observed mass spectra in PDG \cite{pdg1}. Analyzing both types of Regge-like trajectories can, then, derive a reliable prediction for the mass spectra of higher spin-$S$ tensor meson resonances. 

This paper is organized as follows: Sect. \ref{sec2} briefly introduces the hard wall AdS/QCD model,  presenting the spin-$S$ tensor meson resonances and the obtention of their mass spectra. The CE is then calculated for the tensor meson resonances. Hence, two types of configurational-entropic Regge trajectories are interpolated, relating the CE to both the $S$ spin and the tensor mesons mass spectra, being the mass spectra of higher spin-$S$ tensor mesons inferred. The mass spectra of the subsequent members of the tensor mesons family is computed, based upon the experimental mass spectra of the already existing members in PDG  \cite{pdg1}. Sect. \ref{sec3} discusses the results and main conclusions. 

\section{Tensor mesons in the hard wall AdS/QCD and configurational entropy}
The hard wall AdS/QCD model is employed to \label{sec2}
derive the tensor mesons family mass spectra. The gauge/gravity duality between a 4D CFT, on the AdS bulk boundary, and 5D gravity on AdS asserts, in the QCD case, that the fifth dimension along the bulk, $x^5\equiv z$, corresponds to an energy scale in the 4D boundary theory \cite{Karch:2006pv}. Conformal coordinates make the AdS bulk metric to read 
\begin{equation}\label{metrica1}
  d s^2 = \frac{1}{z^2} (d x_{\mu}\,dx^\mu - d z^2)
\end{equation}
Although almost conformal in the UV, small $z$ limit, QCD is a strongly coupled theory that breaks  conformal symmetry in the IR, large $z$, regime. Therefore, a bulk cut-off at $z = z_0$ is needed to preclude this ill-defined region $0\leq z \leq z_0$. 
Here $z_0\sim 1/\Lambda_{\rm QCD}$, meaning that the $\Lambda_{\rm QCD}$ coupling constant decreases as the energy scale increases. Operators in QCD  correspond to fields in the AdS  
bulk \cite{Maldacena:1997re,Witten:1998qj,gub}. 

Taking into account the  tensor meson resonances, the first KK excitation corresponds to the spin-1 $\rho(770)$ meson. This particle, in the hard wall AdS/QCD, is related to a bulk gauge field $V_A$, whose action $S=\int\frac{1}{4}\sqrt{g} F_{AB} F^{AB}{\rm d}^5x $ reads \cite{Katz:2005ir} 
\begin{equation}\label{s1}
  S \!=\! \int\!
   \frac{1}{z} \left(\frac{1}{4} F_{\mu \nu}F^{\mu\nu} - \frac{1}{2} F_{\mu 5}F^{\mu 5} \right){\rm d}^5 x.
\end{equation}
The $\rho(770)$ mesonic state can be modelled by the 4D projection 
$V_{\mu}(x^\mu, z)$ of the bulk gauge field. The other KK resonances, $\rho_n$, are boundary fields whose bulk profiles are governed by the following EOM in the AdS bulk,
\begin{equation}
  \left(\partial_z^2 -\frac{1}{z}\partial_z +  m_n ^2\right) v_n ( z ) = 0,
\label{S1}
\end{equation} with solution 
$  v_n ( z ) = A_n z \left( {}{J}_1 (m_n z) + \alpha_n {}{Y}_1 (m_n z)  \right),$ where $J_1$ and $N_1$ are Bessel functions. 
Neumann and Dirichlet conditions $\rho'(z_0) =0=\rho(0)$ yield $\alpha_n = 0$, thus quantizing the $\rho$ meson family mass spectra by 
${}{J}_0 (m_n z_0) = 0$. Normalizing the $\rho$ function as 
 $\int  \frac{\rho^2_n (z)}{z}\,{\rm d}z = 1$ yields 
\begin{equation}\label{mm1}
  v(z) =  \frac{2.72\,z}{z_0}{J}_1\left(  \frac{2.40 \,z}{z_0}\right)
\end{equation}
The experimental $\rho(770)$ mass, 775.49 MeV, yields  $z_0 = 3.105\times 10^{-3}\, \mathrm{MeV}^{- 1}$.

The next member of the tensor meson family is the $f_2(1270)$ particle resonance. 
To describe it, naturally a  spin-2 bulk field  $h_{\mathrm{AB}}$ must introduced, whose 
lightest KK mode correspond to the  $f_2(1270)$. A candidate for such a bulk spin-2 field is the graviton \cite{Katz:2005ir}. Expanding the AdS background
\begin{equation}
  d s^2 = \frac{1}{z^2}\left(( \eta_{\mu \nu} + h_{\mu \nu} ) d x^{\mu} d x^{\nu}-d z^2\right) 
\end{equation}
yields the bulk  Einstein-Hilbert action to read 
\begin{equation}
-  \int_{\rm AdS}  \frac{1}{2z^3} \left(  h_{\mu \nu} \Box h^{\mu \nu} -\partial_z
  h_{\mu \nu} \partial_z h^{\mu \nu} + \cdots
  \right){\rm d}^5 x. \label{haction}
\end{equation}
Each field has a tower of KK modes, determined by
the EOM 
\begin{equation}
\left(\partial_z^2 -\frac{3}{z} \partial_z  + m_n^2 \right)g_n (z) =0, \label{S2}
\end{equation}
whose solution reads 
\begin{equation}
  g_n (z) = A_n z^2 \left({J}_2 (m_n z) + \alpha_n {Y}_2
  (m_n z)\right).
\end{equation}
Neumann and Dirichlet $g'(z_0)=0=g(0)$ yield $\alpha_n = 0$, being the
masses quantized by the zeros of the equation ${J}_1 (m_n z_0) = 0$. Upon normalization,  $\int_0^\infty \frac{g_n^2}{z^3}\,{\rm d}z = 1$, the $f_2$ tensor meson 
can be described by
\begin{equation}
  g(z) =  \frac{3.51\,z^2}{z_0} {}{J}_2 \left(\frac{3.83\,z}{z_0}\right). \label{hwave}
\end{equation}
Hence, the lightest $f_2$ particle resonance mass is given by $3.83 z_0^{- 1} = 1236$ MeV.
The experimental mass of $f_2(1270)$ in PDG is $1275.1\pm1.2$ MeV \cite{pdg1}.  

In addition to the $f_2(1270)$, higher spin-$S$ resonances constitute the lightest tensor meson family. The hard wall AdS/QCD can derive their masses with a good accuracy up to  spin $S=5$, however for $S\geq 6$ there is already a gap of 14.6\% between the predicted and the experimental mass of the $f_6(2510)$ tensor meson resonance. To derive a more reliable mass spectra for the subsequent members in the tensor meson family, information theory and AdS/QCD will be put together. The configurational entropy will play an important role, together with the experimental masses of the already observed tensor mesons in PDG \cite{pdg1}. Before it, high spin-$S$ mesons must be briefly introduced \cite{Katz:2005ir,Karch:2006pv}.

A KK-type splitting of higher spin-$S$, with $S\geq 3$, string modes are described by the rank-$S$ tensor fields,  
$\psi_{N_1\ldots N_S}$, corresponding to a spin-$S$ state in QCD. Gauge invariance
$\delta \psi_{N_1\ldots N_S}=\nabla_{({N_1}}\zeta_{N_2\ldots N_S)}$ and symmetry justification  are usually used \cite{Katz:2005ir} to derive EOMs for 
the spin-$S$ resonances. In fact, the existence of kinetic terms of higher spin-$S$ fields in AdS, with the above gauge invariance yields the  axial gauge choice $\psi_{5 N_2 \ldots N_S}=0$ \cite{Karch:2006pv}.
This gauge is preserved under $\zeta_{\mu_2 \ldots \mu_S}(x^\mu,z) = z^{2(1-S)}\zeta_{\mu_2 \ldots \mu_S}(x^\mu)$. Hence, 
a field $\psi(x^\mu,z) = z^{2(1-S)} \mathring{\psi} (x^\mu)$ 
can represent a zero mode. 
With respect to $\mathring\psi$, the Lagrangian contains a  kinetic term 
$ z^{1-2S} \partial_\mu\mathring{\psi} \partial^\mu\mathring{\psi}$. 
Hence, the KK modes for spin-$S$ tensor mesons satisfy \cite{Katz:2005ir}
\begin{equation}
\partial_z {z^{1-2S}} \partial_z \mathring{\psi} +{m^2}{z^{1-2S}} \mathring{\psi} =0,
\label{higherS}
\end{equation}
then generalizing Eqs. (\ref{S1}, \ref{S2}) and leading to the known results 
for spin-1 and spin-2 modes.

The normalizable solutions of Eq. \eqref{higherS} are of the form $z^S{}{J}_S(mz)$.  When Neumann 
boundary conditions are imposed in the IR regime, Ref. \cite{Katz:2005ir} derives the mass of the lightest
spin-$S$ particle resonance as the first zero of the ${}{J}_{S - 1}(mz)$ function.
This leads to the tensor mesons spin-$S$ resonances in QCD. 
The first column in Table \ref{scalarmasses} encompasses the spin, $S$, of the respective spin-$S$ tensor meson resonances, whereas the second column depicts the $\rho(770)$, $f_2(1270)$, $\omega_3(1670)$, $f_4(2050)$, $\rho_5(2350)$, $f_6(2510)$ tensor mesons. The third and forth column indicate, respectively, the tensor mesons family mass spectra in  PDG \cite{pdg1} and in AdS/QCD hard wall. {The soft-wall in Ref. \cite{Karch:2006pv}, have the mass spectra $m_{n,S}^2=4k^2(n+S)$, for $k=0.388$ GeV, displayed in the fifth column of  Table \ref{scalarmasses}. This value of $k$ is optimal in what concerns fixing the $\rho(770)$ mass and is most usually employed in the literature.}
\begin{table}[h]
\begin{center}--------- Tensor mesons mass spectra (MeV) ---------\medbreak
\begin{tabular}{||c|c||c|c|c||}
\hline\hline
        $S$ & Tensor mesons       & Experimental & Hard wall&Soft wall  \\\hline\hline
\hline
         \textcolor{black}{1} & $\rho(770)$    &$775.49\pm0.34$     &776  & 776          \\\hline
       \textcolor{black}{2} &   $f_2(1270)$&$1275.1\pm1.2$    &1236&1097\\\hline
         \textcolor{black}{3}& $\omega_3(1670)$           &$1667\pm4$      &1657       &  1344    \\\hline
         \textcolor{black}{4} & $f_4(2050)$         &$2018\pm11 $             &2058&1552 \\\hline
         \textcolor{black}{5*} & $\rho_5(2350)$           &$2330\pm35$                &2448& 1735  \\\hline
         \textcolor{black}{6*} & $f_6(2510)$           &$2469\pm 29 $    &2830&1901                \\
\hline\hline
\end{tabular}
\caption{Mass spectra for the tensor mesons family resonances. The particles with an asterisk are left out the summary table in PDG \cite{pdg1}. }
 \label{scalarmasses}
\end{center}
\end{table}
One might argue whether the AdS/QCD soft wall model can better fit the experimental masses in Table \ref{scalarmasses}.  However 
the soft wall model poorly emulates the experimental mass spectra of the tensor mesons family, {as seen in the fifth column of  Table \ref{scalarmasses}.} The hard wall 
is much more appropriate for better emulating the mass spectra, as Table \ref{scalarmasses} illustrates. 
As the hard wall AdS/QCD model predicts the $f_2(1270)$ mass that differs 3.1\% from the experimental value, and a difference of 14.6\% between the predicted and the experimental mass of the $f_6(2510)$ resonance, we propose a method based on the configurational entropy, for using the experimental data, involving the tensor mesons family mass spectra
to derive the masses of the next generation of tensor mesons resonances. 

For it, Ref. \cite{Karch:2006pv} analogously derives the EOM (\ref{higherS}) from the Lagrangian \begin{equation}
\mathfrak{L}=e^{2A(z)(S-3)} D^N\mathring{\psi}_{\mu_1\ldots \mu_S}D_N\mathring{\psi}^{\mu_1\ldots \mu_S}\label{lagg},\end{equation} for $A(z)=-\log(z)$. Specifically to the tensor meson family, the main tool for computing the CE is any localized function  that represents the tensor mesons in AdS/QCD, namely, the energy density, $\uprho(z)$.
This can be derived from a Lagrangian density, $\mathfrak{L}$, with stress-momentum tensor components given by  
 \begin{equation}
 \!\!\!\!\!\!\!\!T_{AB}\!=\!  \frac{2}{g^{1/2}}\!\! \left[ \frac{\partial (g^{1/2} \mathfrak{L})}{\partial g^{AB} }\!-\!\partial_{ {x}^C }  \frac{\partial (g^{1/2}  \mathfrak{L})}{\partial \left(\!\frac{\partial g^{AB} }{\partial { x}^C}\!\right) }
  \right].
  \label{em1}
 \end{equation} 
 \noindent  
Hence, the energy density is read off the  $T_{00}(z)$ tensor component in Eq. (\ref{em1}). 
The Fourier transform 
$\uprho(k) = \int_0^\infty\uprho(z)e^{-ik z}\,{\rm d}z,$ with respect to the AdS bulk dimension, defines the normalized modal fraction    
\cite{Gleiser:2018kbq,Gleiser:2014ipa,Sowinski:2015cfa}
\begin{eqnarray}
\uprho_{\rm norm}(k) = \frac{|\uprho({k})|^{2}}{\int_0^\infty  |\uprho({k})|^{2}{\rm d}k}.\label{modalf}
\end{eqnarray} Therefore the CE is defined as \cite{Gleiser:2011di,Gleiser:2012tu,Gleiser:2015rwa}
\begin{eqnarray}
{\rm CE}[\uprho] = - \int_0^\infty{{\mathring\uprho_{\rm norm}}}({k})\log {{\mathring\uprho_{\rm norm}}}({k})\, {\rm d}k\,,
\label{confige}
\end{eqnarray}
for ${\mathring\uprho_{\rm norm}}(k)=\uprho_{\rm norm}(k)/({\uprho_{\rm norm}})_{\rm max}(k)$.

According to this procedure, the CE is, then, computed by Eq. (\ref{confige}) for  the lightest family of tensor meson resonances, $\rho(770)$, $f_2(1270)$, $\omega_3(1670)$, $f_4(2050)$, $\rho_5(2350)$, $f_6(2510)$   \cite{pdg1}, and subsequent elements in this family. Firstly, the CE is here computed with dependence on the $S$ spin of the tensor mesons, showed in Table \ref{CES}. Let one denotes by $X_7$ ,  $X_8$, and $X_9$ the next members of the tensor meson family in Table \ref{CES}, 
with spin $S=7$, $S=8$, and $S=9$, respectively.
\begin{table}[h]
\begin{center}\medbreak
\begin{tabular}{||cc||c||c||c||c||c||}
\hline\hline
  &   $S$ & tensor meson&~CE \\\hline\hline
     &  \, 1 \,&\, $\rho(770)$&$3.0467$ \\\hline
     \,&\,   2 \,&\, $f_2(1270)$&$3.1073$ \\\hline
     \,&\,   3 \,&\, $\omega_3(1670)$&$3.1539$\\\hline
     \,&   4\, &\, $f_4(2050)$&$3.1931$ \\\hline
     \,&\,   5 \,&\, $\rho_5(2350)$&$3.2279$\\\hline
     \,&\,   6 \,&\, $f_6(2510)$&$3.2589$ \\\hline
     \,&\,  7 \,&\, $X_7$\,&$3.2765$  \\\hline
     \,&\,8\,&$\,X_8$&$\,3.2936$\\\hline
     \,&\,9\,&$\,X_9$&$\,3.3201$\\\hline
\hline\hline
\end{tabular}\caption{The CE for the tensor mesons family as a function of their spin $S$, in the AdS/QCD model. }
\label{CES}
\end{center}
\end{table}
\begin{figure}[H]
\centering
\includegraphics[width=7.5cm]{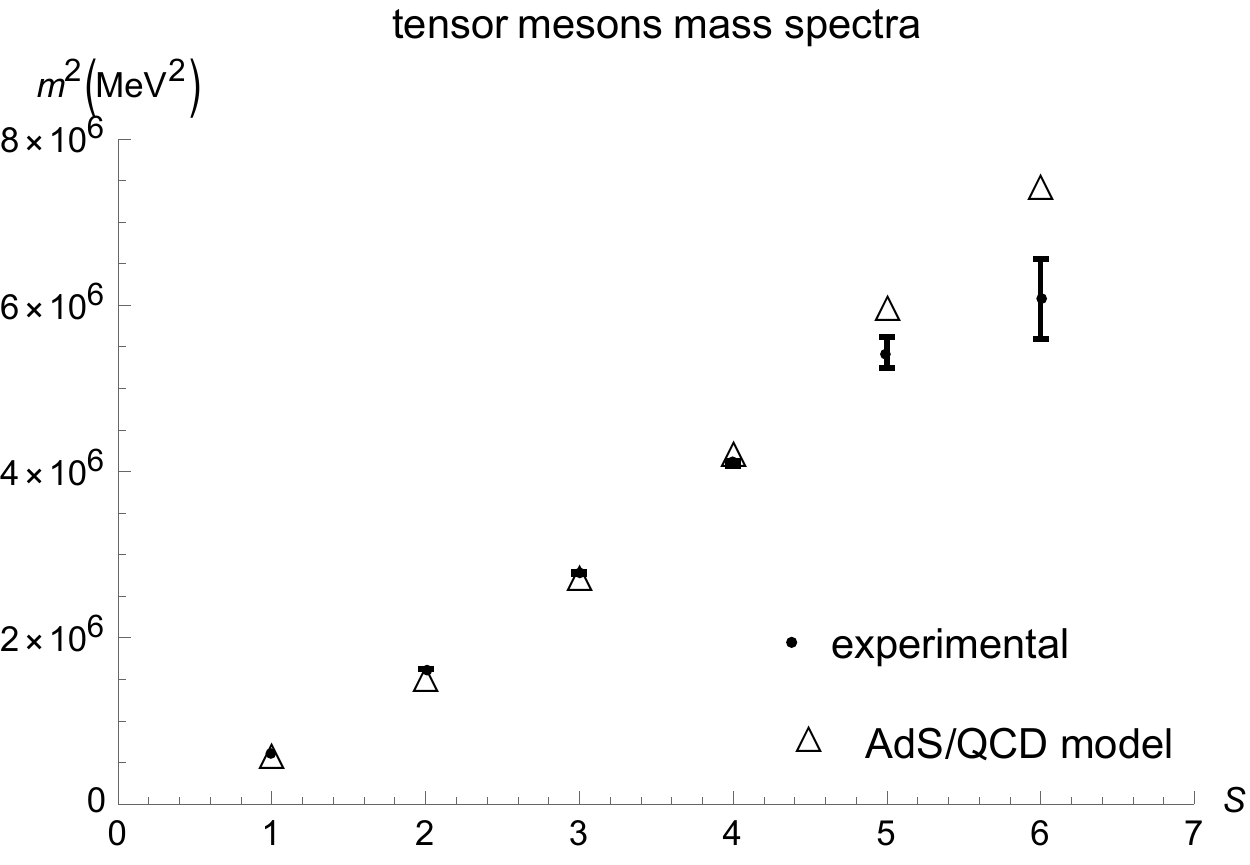}
\caption{Mass spectra of the $\rho(770)$, $f_2(1270)$, $\omega_3(1670)$, $f_4(2050)$, $\rho_5(2350)$, $f_6(2510)$ tensor meson resonances, as a function of their $S$ spin.}
\label{kaon1}
\end{figure}

The data in Table \ref{CES} is better represented by the points in Fig. \ref{kaon1} as a function of their $S$ spin. Besides, interpolation of this data makes possible to introduce a configurational-entropic Regge trajectory, for the tensor mesons resonances, as a dashed curve in Fig. \ref{kaon1}.
\begin{figure}[H]
\centering
\includegraphics[width=7cm]{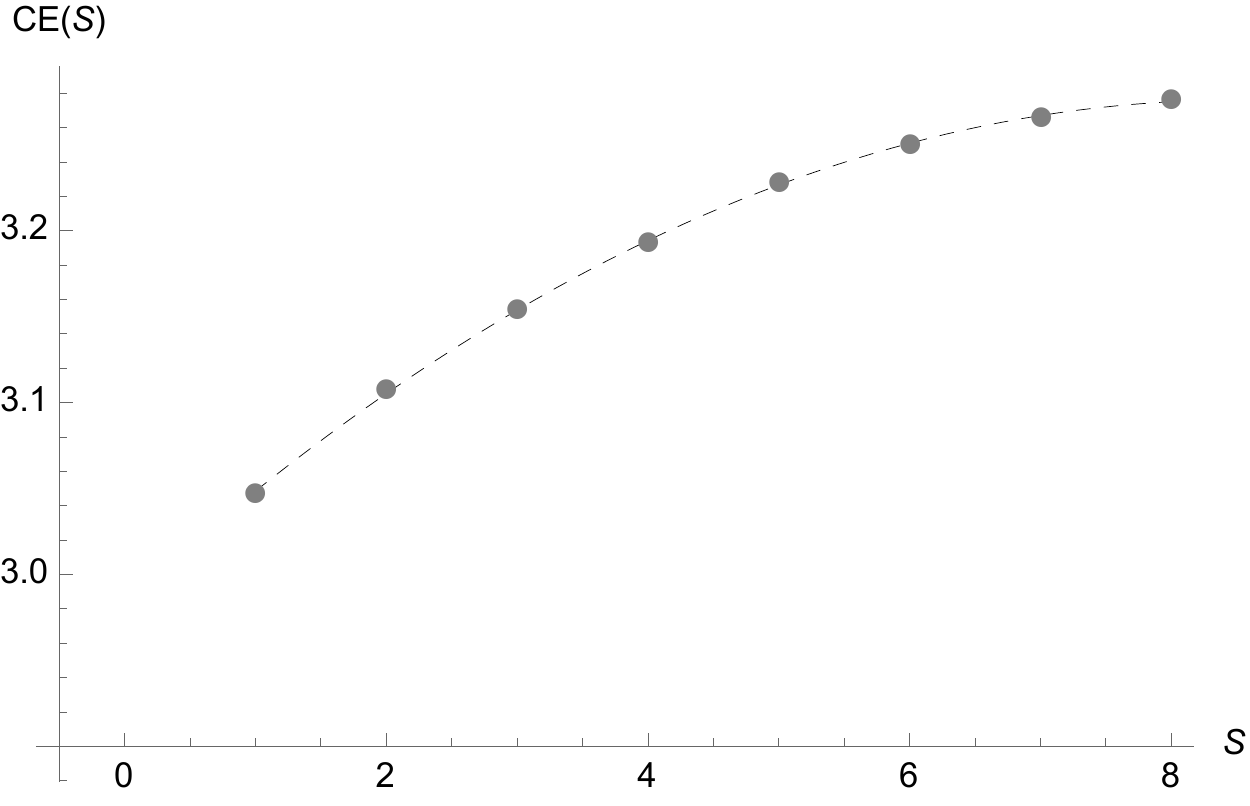}
\caption{CE of the tensor family as a function of the $S$ spin.}
\label{kaon2}
\end{figure}
\noindent The explicit expression of the configurational-entropic Regge trajectory reads
\begin{eqnarray}
 {\rm CE}(S) &=& -0.00343\, S^2+ 0.06572 \,S+ 2.98663.\label{itp1}
   \end{eqnarray} We opted to use a quadratic interpolation, as it is sufficient to delimit within $\sim0.38\%$ the standard deviation. 
    
\begin{figure}[H]
\centering
\includegraphics[width=8cm]{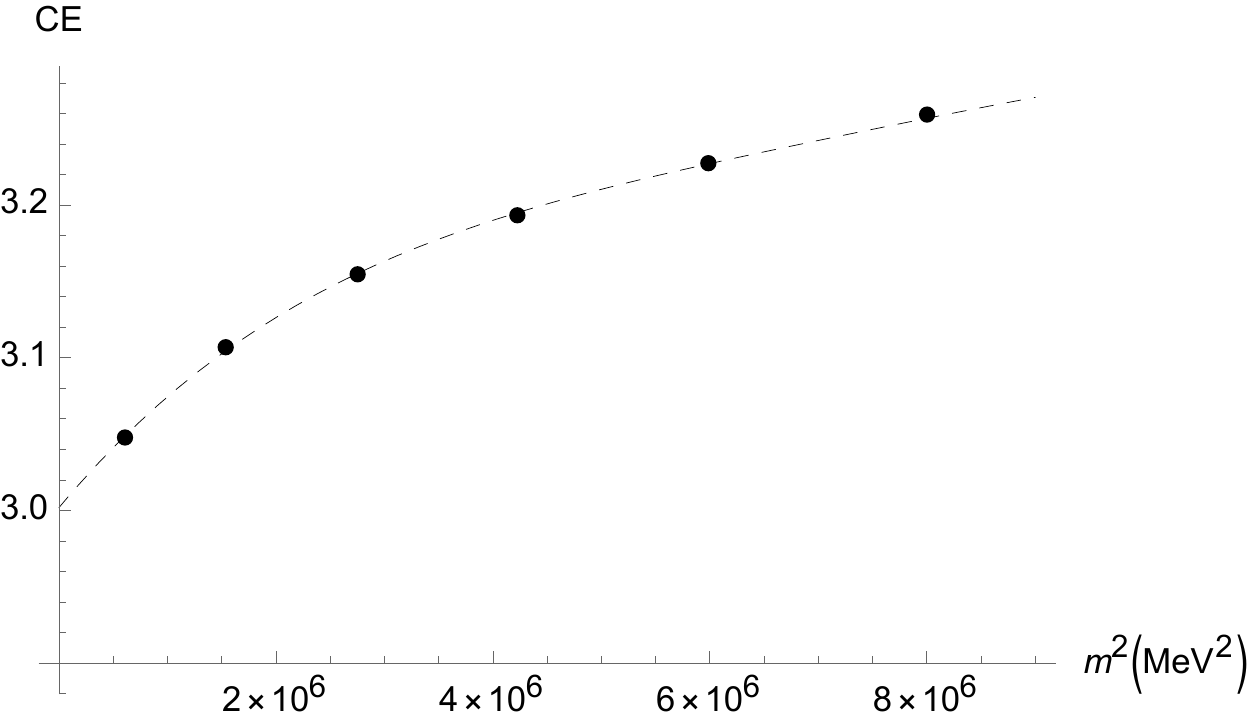}\medbreak
\caption{CE for the $\rho(770)$, $f_2(1270)$, $\omega_3(1670)$, $f_4(2050)$, $\rho_5(2350)$, $f_6(2510)$ tensor mesons resonances, with respect to the mass spectra.}\label{f6}
\end{figure}

 A second type of configurational-entropic Regge trajectory, relating the CE to the tensor meson family (squared) mass spectra, is shown in Fig. \ref{f6}, strictly for the tensor meson resonances in \cite{pdg1}, corresponding to $S\leq 6$. 
 \noindent {\color{black}{The dashed curves in Figs. \ref{kaon2} and \ref{f6} correspond, respectively, to 
Eqs. (\ref{itp1}) and (\ref{itq1}).}}
 For the plot in Fig.  \ref{f6}, the configurational-entropic Regge trajectory has the following 
form:
\begin{eqnarray}\label{itq1}
\!\!\!\!\!\!\!\! {\rm CE}(m) \!&=\!&\! 8.7553\! \times\! 10^{-22}\, m^6\!-\!1.1970\! \times\! 10^{-14}\, m^4\!\nonumber\\&&+8.1127\! \times\! 10^{-8}\,m^2+3.0024,    \end{eqnarray} within $0.14\%$ standard deviation. {\color{black}{It is worth to emphasize that in Eq. (\ref{itq1}) $m$ refers to the mass.}}

 Alternatively of computing the lightest tensor spin-$S$ meson resonances mass spectra, as solutions read off the equation ${J}_{S - 1}(mz)=0$, one can use Eqs. (\ref{itp1}, \ref{itq1}) together. It provides a better accuracy to experimental data in PDG \cite{pdg1}. Subsequently, the mass spectra of the next members of the tensor mesons family, with $S\geq 7$ are derived.  Using the information content of 
 the tensor meson resonances, it is a more realistic one, as it  employs the experimental mass spectra of the $\rho(770)$, $f_2(1270)$, $\omega_3(1670)$, $f_4(2050)$, $\rho_5(2350)$, $f_6(2510)$ tensor meson resonances \cite{pdg1}. 

 The mass spectra for the $X_7, X_8$ and $X_9$ elements can be easily inferred, by employing Eqs. (\ref{itp1}, \ref{itq1}). In fact, for $S=7$, Eq. (\ref{itp1}) yields 
${\rm CE} = 3.27836$. Then substituting this value in the configurational entropic Regge trajectory (\ref{itq1}), and solving the resulting equation, the solution is the mass $m_{X_7}=2619$ MeV, for the $X_7$ tensor meson. It yields a reliable range $2608\; {\rm MeV}\lesssim m_{X_7}\lesssim 2630\;{\rm MeV}$. Similarly for the next member $X_8$ of the tensor mesons family, substituting $S=8$ into Eq. (\ref{itp1}) implies that the corresponding CE has value $3.29254$. Hence, replacing this value into 
the configurational entropic Regge trajectory (\ref{itq1}) yields $m_{X_8}=2690$ MeV the mass range $2675 \,{\rm MeV}\lesssim m_{X_8}\lesssim 2705\, {\rm MeV}$. Besides, the same can be accomplished for the $X_9$ tensor meson, when Eq. (\ref{itp1}) implies that ${\rm CE}_{X_9} = 3.32011$. When replaced into Eq. (\ref{itp1}), it produces the mass $m_{X_9}=2786$ MeV, in the range $2769\, {\rm MeV}\lesssim m_{X_9}\lesssim 3003\, {\rm MeV}$. 

\section{Concluding remarks}\label{sec3}
Interplaying the  CE and the hard wall AdS/QCD model made it possible to derive 
 configurational-entropic Regge trajectories, for the tensor mesons resonances. 
 It was reasonable to consider tensor meson resonances up to spin $S=9$. 
 Two types of Regge-like trajectories were derived, relating the tensor meson family CE to both their spin-$S$ and their mass spectra as well.  Analyzing both types of Regge-like trajectories predicted the mass spectra of high spin-$S$ tensor meson resonances, for $S\geq 7$. Their masses are  respectively given by $m_{X_7}=2619\pm 11$ MeV, $m_{X_8}=2690\pm 15$ MeV and $m_{X_9}=2786\pm 19$. 
These results open two possibilities. Either the $X_7, X_8$ and $X_9$ tensor meson resonances are potentially new particles to be still detected or there exists already detected particles, in the PDG list of particles.  In fact, trying to match the further members of the tensor meson family, there is the  $X(2632)$ element in PDG \cite{pdg1}, with mass $m=2635.2\pm 3.3$, with until unknown spin, that might 
be identified to the $X_7$ member of the tensor meson family. In addition, the element $X(2680)$ in PDG, with mass $m=2676\pm 27$, also having unknown spin, might correspond to the $X_8$ member of the tensor meson family. Analogously, $X(2750)$, with mass $m=2747\pm 32$, might speculatively be the $X_9$ member of the tensor meson family, although this last possibility is less feasible, as PDG lists tensor mesons such that $S\leq 7$, up to the 2018 PDG edition \cite{pdg1}. 
{\color{black}{Finally, our method derived the masses $m_{X_7}=2619\pm 11$ MeV, $m_{X_8}=2690\pm 15$ MeV and $m_{X_9}=2786\pm 19$, whereas the hard wall AdS/QCD model predicts 
 $m_{X_7}= 3206.5$ MeV, $m_{X_8}= 3577.7$ MeV and $m_{X_9}= 3945.2$ MeV.  One may assert that as already the $f_6(2510)$ element, corresponding to ($S=6$), already differs from the value in the PDG by 14.6\%, higher spin tensor meson resonances in the hard wall are not quite reliable for $S>6$.  One can note that this result for the hard wall model disagrees with the expected behaviour of the Regge trajectory for  higher spin-$S$ tensor mesons resonances, for $S>6$. Therefore the procedure using the CE is more trustworthy. }}

\medbreak
\paragraph*{Acknowledgments:}   L. F.  is supported by the National Council for Scientific and Technological Development  -- CNPq (Brazil) under Grant No. 153337/2018-4. RdR~is grateful to FAPESP (Grant No.  2017/18897-8), to  CNPq (Grants No. 406134/2018-9 and No. 303293/2015-2) and  to HECAP -- ICTP, Trieste, for partial financial support, and this last one also for the hospitality.  

\end{document}